\documentclass[proceedings]{JHEP37}

\usepackage{graphicx}
\usepackage{dcolumn}
\usepackage{bm}
\usepackage{xcolor}
\usepackage{eurosym}
\usepackage{amssymb}
\usepackage{amsfonts,cite}
\usepackage{amsmath}
\usepackage{graphicx}
\usepackage{multirow}
\usepackage{epsfig}
\usepackage[utf8]{inputenc}
\usepackage{multicol}
\usepackage{graphicx}
\usepackage{wrapfig}
\usepackage{placeins}
\usepackage{slashed}
\usepackage{lipsum}
\usepackage{array}
\usepackage[thinlines]{easytable}
\usepackage{mathrsfs}
\usepackage[export]{adjustbox}
\usepackage{blindtext}

\title{Phase transitions and thermodynamic cycles in the broken PT-regime}
\author{Andreas Fring$^\bullet$ and Marta Reboiro$^\circ$\\
	$^\bullet$ Department of Mathematics, City, University of London, Northampton Square,\\ $\quad$London EC1V 0HB, UK \\
	$^\circ$Institute of Physics of La Plata (IFLP), Boulevard 113,  La Plata C.P. 1900, Argentina \\
  $\quad$	a.fring@city.ac.uk, reboiro@fisica.unlp.edu.ar}

\conference{Phase transitions and thermodynamic cycles in the broken PT-regime}
\abstract{
We propose a new type of quantum thermodynamic cycle whose efficiency is greater than the one of the classical Carnot cycle for the same conditions for a system when viewed as homogeneous. In our model this type of cycle only exists in the low temperature regime in the spontaneously broken parity-time-reversal (PT) symmetry regime of a non-Hermitian quantum theory and does not manifest in the PT-symmetric regime. We discuss this effect for an ensemble based on a model of a single boson coupled in a non-Hermitian way to a bath of different types of bosons with and without a time-dependent boundary. The cycle can not be set up when considering our system as heterogeneous, i.e. undergoing a first order phase transition. Within that interpretation we find that the entropy is vanishing throughout the spontaneously broken PT-regime.}

\input{tcilatex}
\begin{document}

\section{Introduction}  

Carnot's thermodynamic cycle has been proposed almost 200 years ago in 1824 \cite{carnot2012reflections,pisano2010pr}. According to Carnot's theorem the most efficient engine operates between two heat reservoirs at absolute cold temperature $T_c$ and absolute hot temperature $T_h$ achieving its efficiency at $\eta=1-T_c/T_h$. Here we propose a new type of cycle that has a larger efficiency, for a system when viewed as homogeneous. The proposed cycle bears resemblance to a Stirling cycle, as unlike the Carnot, which moves along two isentropes and two isothermals our cycle moves along two isochorics and two isothermals. We formally identify a combination of coupling constants as the analogue of the volume in this picture. When considering our system as heterogeneous, i.e. undergoing a first order phase transition, the Maxwell construction leads to a breakdown of the features that allow for the set of the proposed cycle. In that interpretation the cycle does not exist.

Our setting is within the context of non-Hermitian ${\cal PT}$-symmetric quantum theories which have been studied extensively for 25 years since their discovery \cite{Bender:1998ke}. Their theoretical description is by now well-understood. {  In contrast with} non-Hermitian open systems, they possess two distinct regimes that are characterised by their symmetry properties with regard to simultaneous parity reflection (${\cal P}$) and time-reversal (${\cal T}$). When their Hamiltonians respect this antilinear symmetry \cite{EW} and their eigenstates are simultaneous eigenstates of the ${\cal PT}$-operator the eigenspectra are guaranteed to be real and the evolution of the system is unitary. This regime is referred to as the ${\cal PT}$-symmetric regime. In turn, when the eigenstates of the ${\cal PT}$-symmetric Hamiltonian are not eigenstates of the ${\cal PT}$-operator, the energy eigenvalues occur in complex conjugate pairs, and one speaks of this parameter regime as the spontaneously broken ${\cal PT}$-regime.  The two regimes are separated in their parameter space by the so-called exceptional point \cite{Kato,berry2004physics,miri2019exceptional}.  Many of the features predicted by these type of theories have been verified experimentally in optical settings that mimic the quantum mechanical description \cite{Guo,ruter2010obs,el2018non}. The transition from one regime to another has recently also been confirmed in a fully fledged quantum experiment \cite{soley2023exp}. 

The new thermodynamic cycle we propose here exists in the spontaneously broken ${\cal PT}$-symmetric regime. In a single particle quantum mechanical theory this parameter regime would normally be discarded on the grounds of being unphysical. The reason for this is that while one of the complex energy eigenvalues will give rise to decay, which is physically acceptable, the other with opposite sign in the imaginary part would inevitably lead to an unphysical infinite growth. One way to fix this problem and ``mend'' the broken regime is to introduce a time-dependent metric \cite{AndTom3} or technically equivalently by introducing time-dependent boundaries \cite{fring2023non}.  In this manner, the instantaneous energy eigenvalues become real in all ${\cal PT}$-regimes. Another possibility is to consider a large thermodynamic ensemble of particles \cite{reboiro2022,ramirez2022psedo}, which is the approach followed here. In that case the average expectation values become real as complex conjugate eigenvalues pair up to make real contributions. We will also explore the combination of both approaches.

\section{A boson coupled to a boson bath}
\subsection{Time-independent scenario}
Our model \cite{fring2019eternal} consists of a boson represented by the operators $a$, $a^\dagger$ coupled to a bath of $N$ different bosons represented by $q_i$, $q_i^\dagger$ $i=1, \ldots , N$. The non-Hermitian Hamiltonian reads
\begin{equation}
H  =  \nu N_a +\nu N_q+ \sqrt{N} (g+k) a^\dag  Q + \sqrt{N} (g-k) Q^\dagger a,
\label{h-nh}
\end{equation}
with number operators $N_a  =  a^\dag a$,  $N_q  =  \sum_{n=0}^N~q_n^\dag q_n$, Weyl algebra generators $Q  =   \sum_n q_n /\sqrt{N}$, $Q\dagger  =   \sum_n q_n^\dagger /\sqrt{N}$ and real coupling constants $\nu$, $g$, $k$. The ${\cal PT}$-symmetry of the Hamiltonian is realised as ${\cal PT}$: $a, a^\dagger, q_i , q_i^\dagger$ $\rightarrow$ $-a,- a^\dagger, -q_i , -q_i^\dagger$. Since the model is non-Hermitian we need to define a new metric in order to obtain a meaningful quantum mechanical Hilbert space or map it to an isospectral Hermitian counterpart. The latter is achieved by using the Dyson map $\eta= e^{\gamma (N_a -Q^\dagger Q)}$ for the similarity transformation in the time-independent Dyson equation
\begin{equation}
	h = \eta H \eta^{-1} = \nu (N_a + N_q)  + \sqrt{N \lambda} ( a^\dag  Q +  Q^\dagger a),  \label{Dysonequn}
\end{equation}

with $ \lambda:= g^2-k^2$ and $\tanh(2\gamma ) = -k/g$. Clearly for $h$ to be Hermitian we require $\lambda > 0$, which constitutes the ${\cal PT}$-symmetric regime, whereas $\lambda < 0$ is referred to as the spontaneously broken ${\cal PT}$-regime when also the eigenstates of $H$ are no longer eigenstates of the ${\cal PT}$-operator.  This is seen from the change in the Dyson map, with $\gamma \notin \mathbb{R}$, in the relation $\phi = \eta \psi$, where $\phi$ and $\psi$ are the eigenstates of $h$ and $H$, respectively. The exceptional point is located $\lambda=0$ in the parameter space where the stated Dyson map becomes undefined. In order to solve the model we can employ the Tamm-Dancoff  method \cite{okuboTD} by mapping the Hermitian Hamiltonian to
\begin{equation}
	h= W_+ \Gamma^\dagger_+ \Gamma_+ + W_- \Gamma^\dagger_- \Gamma_-
\end{equation}
with
\begin{equation}
	W_\pm := \nu \pm \sqrt{N} \sqrt{\lambda}, \qquad    \Gamma_\pm^\dagger = \frac{1}{\sqrt{2}} \left(  a^\dagger \pm Q^\dagger   \right).  
\end{equation}
The eigensystem of $h$ then consists of two decoupled Hermitian harmonic oscillators 
\begin{equation}
	h   \left|  n_+,n_-    \right\rangle  = \left(  E_{n_+} +  E_{n_-}  \right)   \left|  n_+,n_-    \right\rangle,
\end{equation}
where the eigenvalues and eigenstates are 
\begin{equation}
	E_{n_\pm} =  n_\pm W_\pm  , \quad   \left|  n_+,n_-    \right\rangle  =\frac{1}{ \sqrt{n_+ !  n_- !} }  \Gamma_+^{\dagger n_+}  \Gamma_-^{\dagger n_-}  \left|  0,0    \right\rangle  ,   \label{EVtimeind}
\end{equation}
respectively, and $n_\pm \in \mathbb{N}_0$. In the ${\cal PT}$-symmetric regime we restrict our parameter range here to $\nu \geq \sqrt{N} \sqrt{\lambda}$ in order to ensure the boundedness of the spectrum. The Hermitian Hamiltonian $h$ is equivalent to the Hamiltonian $H$ in equation (\ref{h-nh}) as long as the Dyson map is well-defined.
\subsection{Time-dependent scenario}
Next, we introduce an explicit time-dependence into our system. This can be achieved in two different ways:  by allowing the non-Hermitian Hamiltonians and the metric to be explicitly time-dependent or by allowing only the metric to be time-dependent. An alternative, but equivalent, viewpoint of these settings correspond to restricting the domain of the system by introducing a time-dependent boundary \cite{fring2023non}. While the latter approach is more physically intuitive, dealing with time-dependent Dyson maps or metric operators is technically easier and better defined. In either case, the Dyson equation (\ref{Dysonequn}) has to be replace by its time-dependent version \cite{fringmoussa}
\begin{equation}
	h(t) = \eta(t) H(t) \eta^{-1}(t)  + i \hbar \partial_t \eta(t) \eta^{-1}(t)  , \label{tDysonequn}
\end{equation}
and one needs to distinguish the non-Hermitian Hamiltonian $H(t)$ from the instantaneous energy operator
\begin{equation}
	E(t) = H(t)   + i \hbar \eta^{-1}(t)   \partial_t \eta(t)  .  \label{tenergy}
\end{equation}
Keeping $H$ time-independent, a solution to (\ref{tDysonequn}) for the non-Hermitian Hamiltonian in (\ref{h-nh}) in form of a time-dependent Dyson map 
\begin{equation}
	\eta(t) =   e^{-i \nu t (N_a+N_q)  - i \mu_I(t) ( a^\dag  Q +  Q^\dagger a)  },
\end{equation}
and a time-dependent Hermitian Hamiltonian 
\begin{equation}
	h(t) =  \nu (N_a + N_q) + \mu(t)( a^\dag  Q +  Q^\dagger a),
\end{equation}
were found in \cite{fring2019eternal}, with  
\begin{equation}
	\mu(t) =   \frac{\lambda \sqrt{N}  \sqrt{c_1^2   + \lambda}   }{  2 \lambda  + 2 c_1^2 \sin^2 \left[  2 \sqrt{N  \lambda}  (t+c_2)     \right] },
\end{equation}
and
\begin{equation}
\!\!\!	\mu_I(t) =   \frac{1}{2} \arctan\left\{  \frac{\sqrt{c_1^2 +\lambda  }}{\sqrt{\lambda}}       \tan\left[  2 \sqrt{N \lambda}  (t + c_2)       \right] \right\} .
\end{equation}
We have set $\hbar =1$ with $c_1$ and $c_2$ being real integration constants. The latter may be set to zero as it just corresponds to a shift in time, whereas the appropriate choice of $c_1$ is crucial since it controls in part the reality of the coefficient functions $\mu(t)$ and $\mu_I(t)$.  

As the operator structure of the time-independent and time-dependent system are identical, they also share the same eigenstates, where the instantaneous energy eigenvalues become  
\begin{equation}
  	E_{n_\pm}(t) =  n_\pm W_\pm(t), \quad \text{with} \,\,  W_\pm(t)= \nu \pm \mu(t) . 
  	\label{EVtimedep}
  \end{equation}
  At the particular times
  \begin{equation}
  	t_c^n=  \frac{1}{4
  		\lambda  \sqrt{N}}   \arccos \left[ 1+ \frac{2 \lambda -\sqrt{\lambda } \sqrt{c_1^2+\lambda }}{c_1^2}\right]+\frac{\pi n }{2 \lambda  \sqrt{N}},   \label{timetc}
  \end{equation}
with $n \in \mathbb{Z}$, the time-dependent system coincides with the  time-independent one. These times are real in the two parameter regimes of either ${\cal PT}$-regime, i.e. $- c_1^2/3 \leq \lambda  \leq 0$ or $0  \leq \lambda  \leq  c_1^2/3$.

Here, we will discuss the thermodynamic properties of the time-independent and time-dependent systems in all ${\cal PT}$-regimes, but not in terms of microstates in bipartite systems as previously done in \cite{fring2019eternal,moise2023ent}. Instead, here will look at large ensembles and in particular focus on setting up a Carnot cycle and a new cycle moving along different thermodynamic paths. In general, quantum thermodynamic properties for non-Hermitian systems have been discussed previously in \cite{jakubsky2007th,jones2010quantum,gardas2016non,reboiro2022,ramirez2022psedo} 

\section{Carnot $(TS)$ vs Stirling $(T \lambda)$ cycles}

\begin{figure}[h]
	\centering       
	\noindent	\begin{minipage}[b]{0.49\textwidth}      
		\includegraphics[width=\textwidth]{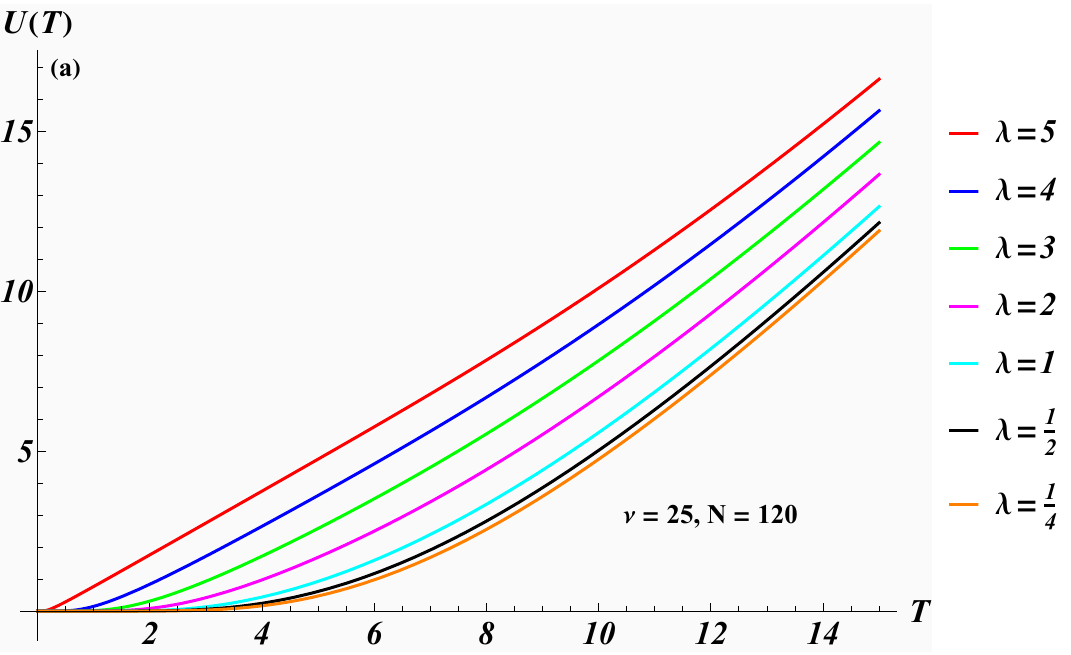}
	\end{minipage}   
	\begin{minipage}[b]{0.49\textwidth}           
		\includegraphics[width=\textwidth]{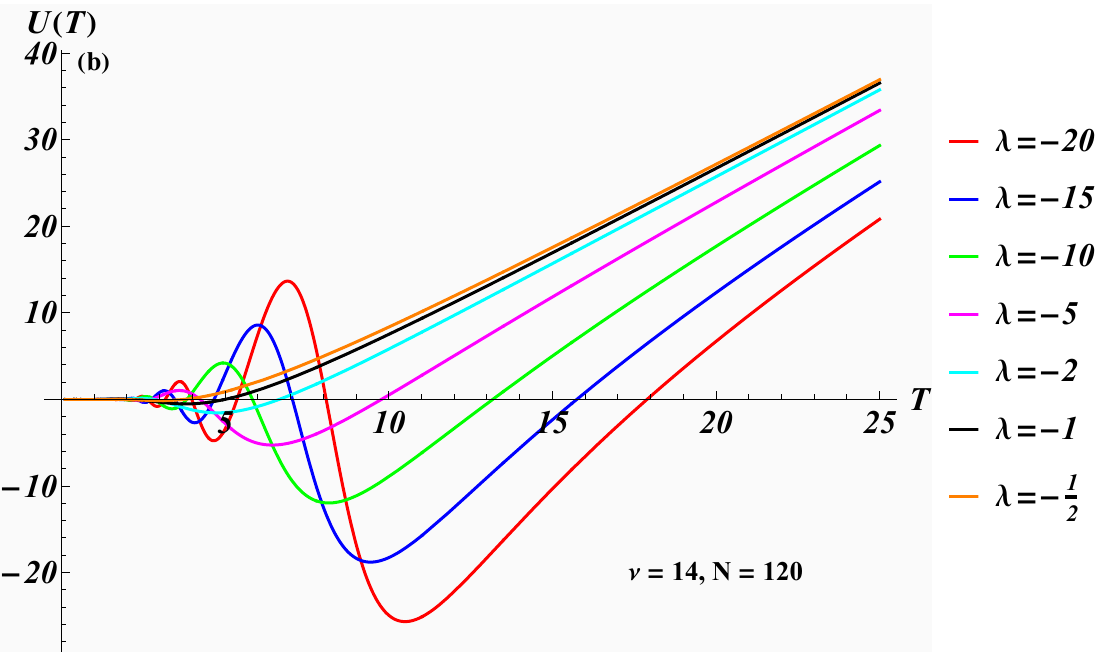}
	\end{minipage}
	\noindent	
	\begin{minipage}[b]{0.49\textwidth}     
		\includegraphics[width=\textwidth]{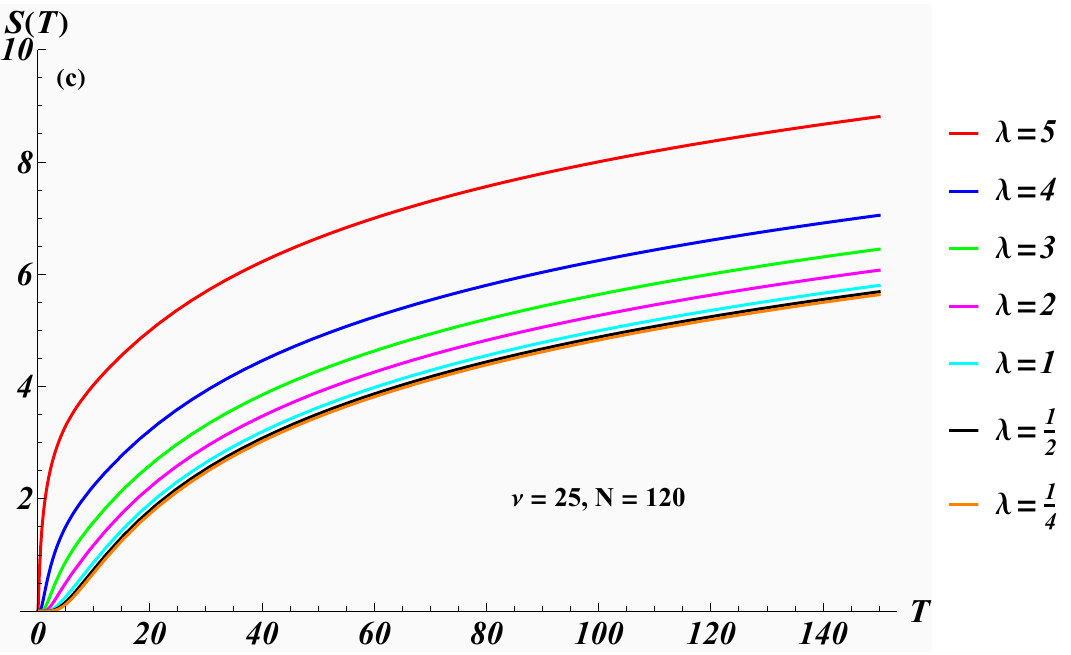}
	\end{minipage}   
	\begin{minipage}[b]{0.49\textwidth}           
		\includegraphics[width=\textwidth]{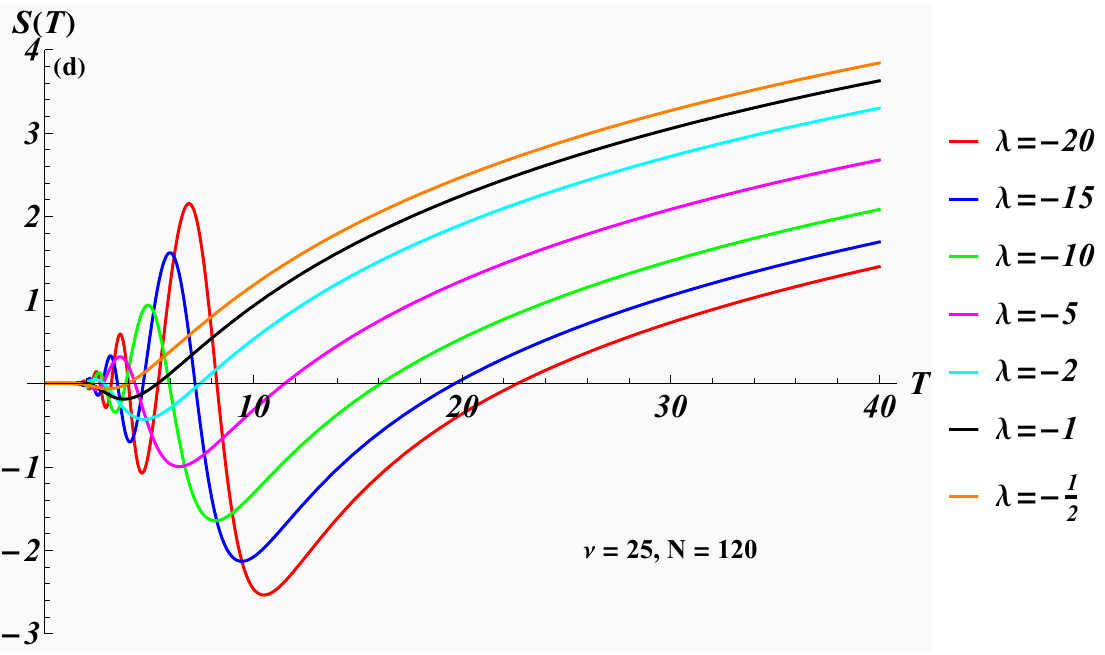}
	\end{minipage} 			
	\caption{Internal energy $U$ {  panels (a), (b) and Entropy panels $S$ (c), (d) as a function of temperature $T$ at different values of $\lambda$ in the $\cal{PT}$-symmetric (left) and in the spontaneously broken $\cal{PT}$-regime.} }
	\label{IntEn}
\end{figure}

The quantum mechanical partition function for canonical ensembles is calculated in a standard fashion for our time-independent model (\ref{h-nh}) as
\begin{eqnarray}
	Z(T,\nu,\lambda) &=& \text{tr}  \rho_h = \sum_{n_\pm }  \left\langle n_+,n_- \right|   \rho_h  \left|  n_+,n_-    \right\rangle \label{partition}  \\
	&=&\frac{ e^{\nu/T}  }{ 4 \sinh( W_+/2T) \sinh( W_-/2T) }  , \notag
\end{eqnarray}
with $\rho_h=e^{-\beta h}$, $\beta=1/T$. From this expression we may compute all thermodynamic quantities that are of interest here. The Helmholtz free energy, internal energy and entropy results in 
\begin{eqnarray}
	F(T,\nu,\lambda) &=&- T \ln Z(T,\nu,\lambda) , \\
	U(T,\nu,\lambda) &=&  \frac{T^2}{Z}  \frac{dZ}{dT} =  \frac{ \text{tr} \left( h  \rho_h  \right)  }{\text{tr}  \rho_h }  \label{Uexp} \\
	&=&\frac{1}{2}    \left[ W_-  \coth \frac{W_-}{2T }+   W_+  \coth \frac{W_-}{2T }  -2 \nu    \right],  \notag  \\
	S(T,\nu,\lambda) &=& \left. - \frac{dF}{dT} \right|_\lambda  =   \ln Z + \frac{U}{T} ,   \label{entropy}
\end{eqnarray}
respectively. The behaviour of these quantities as functions of temperature, displayed in Figure \ref{IntEn}, is qualitatively different in the two $\cal{PT}$-regimes discussed here. 

We find that in the $\cal{PT}$-symmetric regime the internal energy, as well as the entropy, behave in a standard fashion with the latter being a monotonously increasing function. Remarkably the energy has been mended as it is also real in the spontaneously broken $\cal{PT}$-symmetric regime. {  This is due to the fact that the complex energies always come in complex conjugate pairs so that their combined contribution in the ensemble always gives a real contribution.} In the low temperature regimes we observe oscillatory behaviour for both quantities, whereas for large temperatures the asymptotic behaviour is similar to the one in the $\cal{PT}$-symmetric regime, with $\lim_{T \rightarrow  \infty} U(T) \sim 2 T$ and $\lim_{T \rightarrow  \infty} S(T) \sim 2 \ln T$.  

We also notice that while the entropy is positive and strictly increasing in the $\cal{PT}$-symmetric regime, it becomes negative and oscillating in the low temperature regime of the spontaneously broken $\cal{PT}$-regime. Thus the second law of thermodynamics appears to be broken. In section IV we discuss how this may be rectified. 

\begin{figure}[h]
	\centering     
	\begin{minipage}[b]{0.90\textwidth}     
		\includegraphics[width=\textwidth]{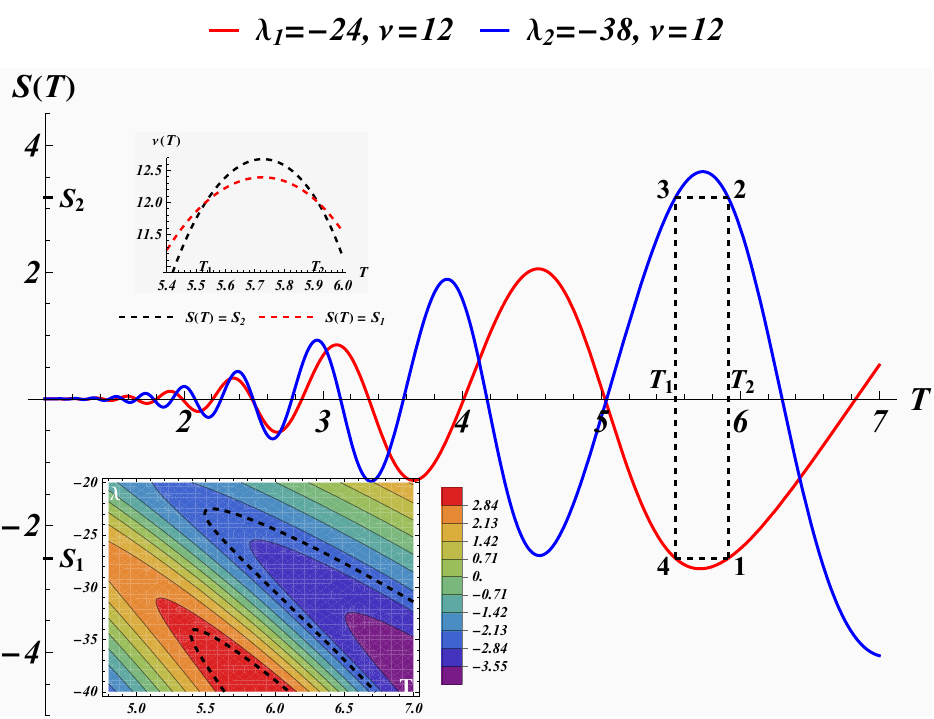}
	\end{minipage}   
		\caption{Entropy as a function of temperature in the spontaneously broken $\cal{PT}$-regime with a Carnot (dashed lines) and new type of thermodynamic cycle. We kept the size of the bath fixed with $N=160$. For the chosen parameters we obtain as solutions of (\ref{condCarnot11}), (\ref{condCarnot}) the temperatures $T_1 = 5.53240$, $T_2 = 5.91528$ and entropies $S_1 = -2.51338$ and $S_2 = 3.16977$. The top inset shows how to vary $\nu$ from $T_2$ to $T_1$ as a function of $T$ and vice versa along constant entropies. The lower inset shows a contour plot of the entropy in the $\lambda$T-plane. } 
		\label{Carnot}
\end{figure}

\begin{figure}[h]
	\centering     
	\begin{minipage}[b]{0.90\textwidth}     
		\includegraphics[width=\textwidth]{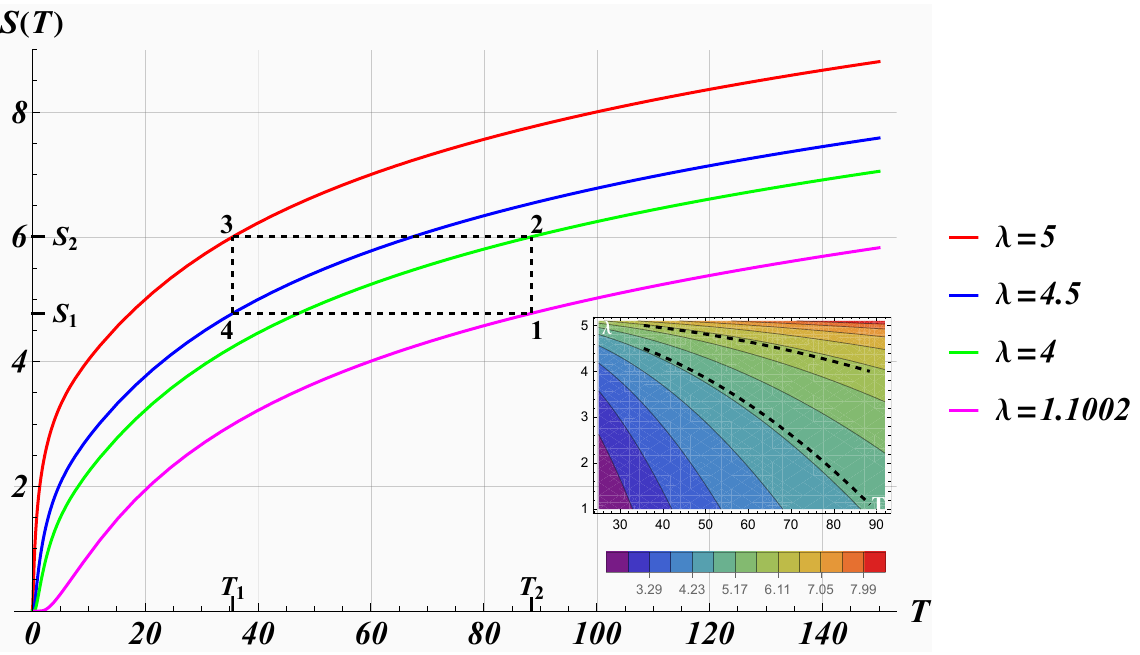}
	\end{minipage}   
		\caption{Entropy as a function of temperature in the $\cal{PT}$-symmetric regime with a Carnot thermodynamic cycle (dashed lines). The size of the bath is $N=120$ and $\nu=25$. The Carnot cycle is constructed between the temperatures $T_1 = 35.5489$, $T_2 = 88.4576$ and entropies $S_1 = 4.7726$ and $S_2 = 6$. The inset shows how to vary $\lambda$ from $T_2$ to $T_1$ as a function of $T$ and vice versa along the constant entropies $S_2$ and $S_1$, respectively. } 
		\label{CarnotPT}
		\end{figure}

We can exploit these features to set up a new type of thermodynamic cycle along a different path and compare with the conventional Carnot cycle in the two ${\cal PT}$-regimes which is identified in Figures \ref{Carnot} and \ref{CarnotPT} in form of a dashed rectangle.  In general, the Carnot cycle is defined as a four step process consisting of an isothermal expansion ($1 \rightarrow 2$), an isentropic expansion ($2 \rightarrow 3$), a subsequent isothermal compression ($3 \rightarrow 4$) and an isentropic compression ($4 \rightarrow 1$).

In our example in the spontaneously broken $\cal{PT}$-regime these steps can be realised by a suitable tuning of the parameters at our disposal. We have:  step $1 \rightarrow 2:$ change $\lambda_1$ to $\lambda_2$,  step $2 \rightarrow 3:$ change $\nu$ as a function of $T$ along the line $S(T) = S_2$ for constant $\lambda$ as indicated in the top inset of Figure \ref{Carnot}, step $3 \rightarrow 4:$ change $\lambda_2$ to $\lambda_1$ and finally in step $4 \rightarrow 1:$ change $\nu$ as a function of $T$ along the line $S(T) = S_1$ for constant $\lambda$ as indicated in the top inset of Figure \ref{Carnot}.

Notice that the steps $2 \rightarrow 3$ and $4 \rightarrow 1$ along the isentropes can not be realised by varying $\lambda$ as a function of $T$ for constant $\nu$. The multi-valuedness of $\lambda (T)$ which makes this impossible can be seen in the contour plot in the lower inset in Figure \ref{Carnot}. However, in the broken $\cal{PT}$-symmetric regime we also have a second option at our disposal to connect the point $2$ with $3$ and $4$ with  $1$. Instead of keeping $\lambda$ fixed and varying $\nu$ along the isentropic, we can keep both $\lambda$ and $\nu$ fixed with only varying the temperature, i.e. we connect the points along the iso-$\lambda$ and iso-$\nu$ lines. This is the new thermodynamic cycle we propose.

The cycle can almost be interpreted as an analogue to the Stirling cycle: Seeking out conjugate pairs of variables in parameter space we may interpret $\lambda$ as the volume and pair it as usual with the pressure $p$. Keeping $\nu$ constant, the total differential of the Helmholtz free energy then acquires the form
\begin{equation}
dF = - S dT - p d \lambda   ,
\end{equation}
such that
\begin{equation}
\!\!	p = \left. - \frac{\partial F}{\partial \lambda} \right|_T=  \frac{\sqrt{N} \sinh \left(\frac{\sqrt{\lambda }
			\sqrt{N}}{T}\right)}{\sqrt{\lambda } \left[2 \cosh \left(\frac{\nu
		}{T}\right)-2 \cosh \left(\frac{\sqrt{\lambda }
			\sqrt{N}}{T}\right)\right]}.    \label{pressure}
\end{equation}

We can then interpret the thermodynamic processes in the new cycle as: $1 \rightarrow 2$: isothermal heat addition, $2 \rightarrow 3$: isochoric (iso-$\lambda$) heat {  addition}, $3 \rightarrow 4$: isothermal heat removal and $4 \rightarrow 1$: isochoric (iso-$\lambda$) heat addition. {  Notice that our cycle differs from a standard Stirling cycle in step $2 \rightarrow 3$, where instead of removing heat we are adding heat.}

As seen in Figure \ref{IntEn} panels (c) and (d), it is crucial to note that in the $\cal{PT}$-symmetric regime and the high temperature regime of the spontaneously broken $\cal{PT}$- regime two points with the same entropy always have different values for $\lambda$ when $\nu$ is fixed or vice versa, since the entropy is monotonously increasing. Hence, the proposed cycle can not manifest in those regimes.

A necessary condition for the cycle, as depicted in Figure \ref{Carnot}, to manifest, is the existence of solutions to the two equations 
\begin{eqnarray}
	S(T_1,\nu,\lambda_1,N) &=&S(T_2,\nu,\lambda_1,N)=S_1,   \label{condCarnot11} \\
	S(T_1,\nu,\lambda_2,N) &=&S(T_2,\nu,\lambda_2,N)=S_2 \label{condCarnot}
\end{eqnarray}
for $T_1$ and $T_2$ with given $N, \nu, \lambda_1,\lambda_2$. Our numerical solutions for these equations are reported in the captions of Figure \ref{Carnot}. Notice that it is by no means guaranteed that for a given set of parameters real solutions to (\ref{condCarnot11}) and (\ref{condCarnot}) exist and that even an ideal Carnot cycle can be realised. In the $\cal{PT}$-symmetric regime no such solution exists. The fact that we found a solution to vary along the isentropics with a single parameter, i.e. $\nu$, is also not guaranteed in all parameter settings.

The newly proposed cycle does indeed beat the Carnot cycle in the sense that the amount of total energy transferred as work $W$ during the cycle as well as its efficiency are larger than in the Carnot cycle. To see that we calculate the work $\Delta W_{ij}$ as the heat $\Delta Q_{ij}$ transferred minus the internal energy $\Delta U_{ij}$ for each of the steps $i \rightarrow j$
\begin{equation}
	                \Delta W_{ij}  =  \Delta Q_{ij} -   \Delta U_{ij}  ,   \label{WQU}
\end{equation}	
where $  \Delta Q_{ij} = \int_{S_i}^{S_j}  T dS $,  $  \Delta W_{ij} = \int_{\lambda_i}^{\lambda_j}  p d \lambda $, with the pressure $p$ identified in (\ref{pressure}), and $\Delta U_{ij} = U_j -U_i$. This means we are adopting here the conventions $ \Delta Q>0$ ($ \Delta Q<0$) for heat absorbed (released) by the system and $ \Delta W>0$ ($ \Delta W<0$) for work done by (put into) the system. The numerical values for our example are reported in \ref{table1}: 
  \begin{table}
  	\centering  
  	\begin{tabular}{ c | |  r |  r | r }
  		$T \lambda$-cycle \,\,& $ \,\,\Delta W_{ij} \,\,$    & $  \,\, \Delta Q_{ij}  \,\,$   &  $  \,\, \Delta U_{ij}  \,\,$  \\  \hline \hline
  		$1 \rightarrow 2 $ & 2.1172 & 33.6174    &  31.5002    \\  
  	$2 \rightarrow 3 $ & 0 &    0.1054   &  0.1054     \\
  	$3 \rightarrow 4 $ &  0.2065      &   -31.4415   &  -31.6480    \\
  	$4 \rightarrow 1 $ &  0      &  0.0424   &  0.0424    \\
  	$	\oint_{\Gamma_2}$ &  2.3238     &  2.3238   & 0     \\
  	\end{tabular}
  	\caption{\label{table1} Contributions to the work $\Delta W_{ij}$, heat $\Delta Q_{ij}$ and internal energy $\Delta U_{ij}$ for different steps $i \rightarrow j$ in the $T \lambda$-cycle.}
  	 \end{table}
  Here each column is computed separately and the assembled results confirm the relation (\ref{WQU}). We obtain the values $U_1= -14.0513$, $U_2= 17.4488$, $U_3= 17.5543$, $U_4= -14.0937$ from (\ref{Uexp}). We denote the path along the dashed rectangle in Figure \ref{Carnot} as $\Gamma_1$ and $\Gamma_2$ as the path that differs from $\Gamma_1$ in the verticals by tracing over the arches above and below the segments $23$ and $41$, respectively. The internal energy is vanishing along any closed loop and therefore does not contribute to the total work. Hence, in our $T \lambda$-cycle the heat is directly converted into work
  \begin{equation}
  	W_{T \lambda}= \oint_{\Gamma_2}  T  dS = 2.3238 .   \label{TL}
  \end{equation}	
  The efficiency, defined in general as the total work done by the system divided by the heat transferred into it, results for our cycle to
  	\begin{equation}
  		\eta_{T \lambda}=\frac{W_{T \lambda}}{\Delta Q_{12}+\Delta Q_{23}+\Delta Q_{41}} = 0.0688. \label{effLT}
  	\end{equation}	
  At first we compare this to the efficiency of the Stirling cycle in an ideal gas
 \begin{equation}
 	\eta_{\text{Stirling}}=\frac{R(T_2-T_1)}{ R T_2 + c_v (T_2-T_1)/ \ln{\lambda_2/\lambda_1} }
 \end{equation}	
  with $R$ denoting the ideal gas constant and $c_v$ the specific heat. With a typical value of $c_v = 5/4 R$ for air this yields $\eta_{\text{Stirling}}= 0.05503$ and if we want to match the expression with $\eta_{T \lambda}$ we would require a negative specific heat of $c_v = -0.4516 R$. Evidently, this means our system is far from an ideal gas.
  
  Next, we compare with the Carnot cycle as indicated in Figure \ref{Carnot}. We report once more the individual contributions in a table:
   \begin{table}
   	\centering  
  	\begin{tabular}{ c | |  r |  r | r }
  		$T S$-cycle \,\,& $ \,\,\Delta W_{ij} \,\,$    & $  \,\, \Delta Q_{ij}  \,\,$   &  $  \,\, \Delta U_{ij}  \,\,$  \\  \hline \hline
  		$1 \rightarrow 2 $ & 2.1172 & 33.6174    &  31.5002    \\  
  		$2 \rightarrow 3 $ & -0.1054 &    0   &  0.1054     \\
  		$3 \rightarrow 4 $ &  0.2065      &   -31.4415   &  -31.6480    \\
  		$4 \rightarrow 1 $ &  -0.0424     &  0   &  0.0424    \\
  		$	\oint_{\Gamma_1}$ &  2.1760     &  2.1760   & 0     \\
  	\end{tabular}
  	{ 	\caption{\label{table2} Contributions to the work $\Delta W_{ij}$, heat $\Delta Q_{ij}$ and internal energy $\Delta U_{ij}$ for different steps $i \rightarrow j$ in the standard Carnot-cycle.}}
  \end{table}
  Thus the total work done by the system is
\begin{eqnarray}
	W_{\text{Carnot}}&=&   	\oint_{\Gamma_1}     dQ  -   \oint_{\Gamma_1}     dU    \\
	&=&	\oint_{\Gamma_1}   T  dS = (T_2-T_1) (S_2-S_1) = 2.1760 ,\notag  
\end{eqnarray}
which is smaller than the work done by the $T \lambda$-cycle (\ref{TL}). The efficiency is obtained in this case as 
 \begin{equation}
	\eta_{\text{Carnot}}= \frac{W_{\text{Carnot}}}{ \Delta_{12}} = 1- \frac{T_1}{T_2} = 0.06473,
\end{equation}	 
which is also smaller than the one obtained for the $T \lambda$-cycle (\ref{effLT}).

In comparison, in the $\cal{PT}$-symmetric regime any Carnot cycle must connect four different values of $\lambda$ or $\nu$ for fixed $\nu$ or $\lambda$, respectively. This is seen in Figure \ref{CarnotPT}  for the first case. A similar Figure can be constructed by varying $\nu$ for fixed $\lambda$. Thus, the new cycle we proposed for the  spontaneously broken $\cal{PT}$-regime can not manifest in the $\cal{PT}$-symmetric regime. A further difference is that when $\nu$ is kept fixed we can not vary $\lambda$  along an isentropic line in the spontaneously broken $\cal{PT}$-regime, whereas in the $\cal{PT}$-symmetric regime we have to vary $\lambda$ to stay on the isentropic.    

Next, we carry out a similar analysis for the time-dependent system. The thermodynamic quantities can be computed in almost the same manner as in (\ref{partition})-(\ref{entropy}), with the difference that the time-dependence is introduced by replacing the functions $W_\pm$ in (\ref{EVtimeind}) by their time-dependent versions $W_\pm(t)$ from (\ref{EVtimedep}). For fixed values of time we obtain a similar behaviour as in the time-independent case and as pointed out in (\ref{timetc}), for some values of time this becomes even identical. 

\begin{figure}[h]
	\centering         
	\begin{minipage}[b]{0.49\textwidth}     
		\includegraphics[width=\textwidth]{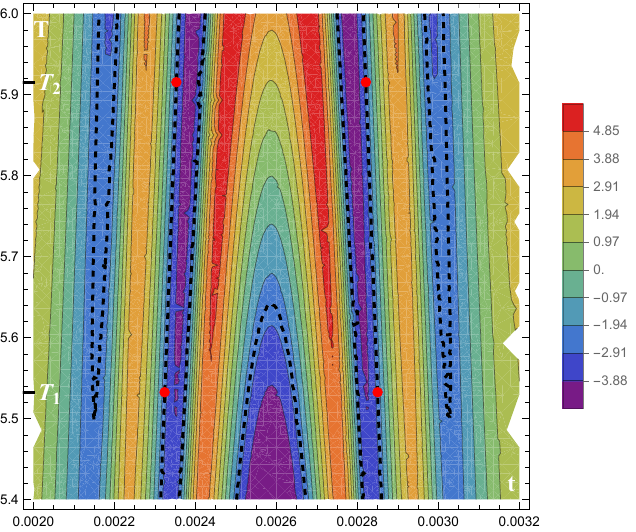}
	\end{minipage}   
	\begin{minipage}[b]{0.49\textwidth}     
		\includegraphics[width=\textwidth]{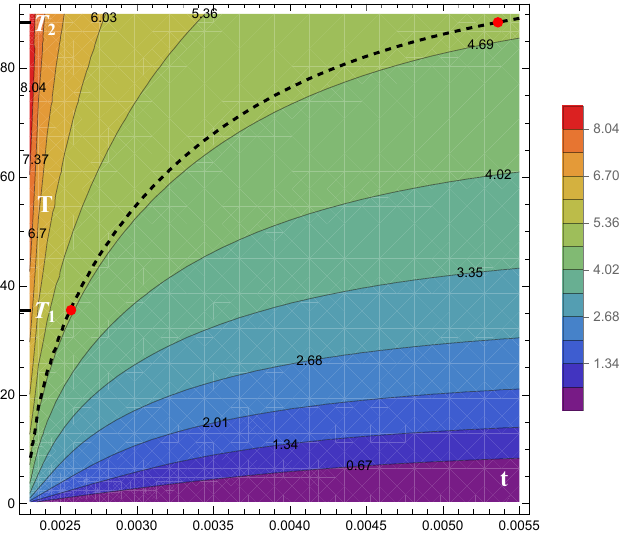}
	\end{minipage} 
	\caption{Panel (a) Contours of constant entropy in the spontaneously broken $\cal{PT}$-symmetric regime in the $Tt$-plane for $N=160$, $\nu=12$, $\lambda = -24$ and $c_1= 4.75$. The dashed lines display the values of constant $S_1$ as specified in Figure \ref{Carnot}. The red dots indicate $S_1=S(T_1,t_1)=  S(T_2,t_2)$ with $t_1 = 0.0023241$, $t_2= 0.0023532$ or $t'_2= 0.0028210$, $t'_1 =0.0028501 $. Panel (b) Contours of constant entropy in the $\cal{PT}$-symmetric regime in the $Tt$-plane for $N=120$, $\nu=25$, $\lambda = 4.5$ and $c_1= 6$. The dashed lines display the values of constant $S_1$ as specified in Figure \ref{IntEn}. The red dots indicate $S_1=S(T_1,t_1)=  S(T_2,t_2)$ with $t_1= 0.0025630$ and $t_2 = 0.0053601$. } 
	\label{EntropyTtfig}
\end{figure}

The novel option we have in the time-dependent case is that we can keep all the model parameters fixed and let the system evolve with time. An example of such an evolution in the spontaneously broken $\cal{PT}$-symmetric regime is seen in Figure \ref{EntropyTtfig}, where we depict contours of constant entropy in the temperature-time plane. We observe that there exist plenty of timelines along the constant entropy contour $S(T) = S_1$, displayed as dashed black lines. After changing from $\lambda_1$ to $\lambda_2$, a similar Figure can be obtained for $S(T) = S_2$. Thus, for the time-dependent system in the broken regime we may lower or increase the temperature  along the isentropics by letting the system evolve in time, which means there exists yet another possibility to manifest the steps $2 \rightarrow 3$ and $4 \rightarrow 1$ in the Carnot cycle. We note that the timescales involved for this process are extremely short, e.g. for the sample values in Figure \ref{EntropyTtfig} panel (a) we have $\Delta_t :=t_2 -t_1 =t'_1-t'_2  = 0.0000291$.

We compare these finding now with the time evolution in the $\cal{PT}$-symmetric regime. As seen in Figure \ref{EntropyTtfig} panel (b), unlike as in the time-independent regime, we have now the option to connect points at different temperatures for the same value of $\lambda$ along an isentropic. Thus in principle we could modify the Carnot cycle displayed in Figure \ref{CarnotPT} and set it up between just two values of $\lambda$, similar as for the broken regime. However, the time-evolution is always increasing the temperature, which is fine for the $4 \rightarrow 1$ step, but for the step $2 \rightarrow 3$ we need to lower the temperature which would require time to run backwards. Hence, a Carnot cycle between two values of $\lambda$ does not exist in the $\cal{PT}$-symmetric regime.       

\section{First order phase transition}

The observed behaviour in the spontaneously broken $\cal{PT}$ regime suggests that various fundamental principles of thermodynamics are apparently violated. The pressure (\ref{pressure}) exhibits regions in which it increases with volume $\lambda$, thus breaking the condition for stability of a thermodynamic system in equilibrium. Moreover, in figure \ref{IntEn} we observe that the entropy does not only take on negative values, but may even decrease so that the second law of thermodynamics is broken. It is these features that allow to set up the proposed cycle.

We may overcome the aforementioned breaches by viewing the system of being in a heterogeneous rather than a homogeneous phase.  Insight into the existence and stability properties of homogeneous and heterogenous states can be obtained from the Maxwell construction and subsequent spinodal decomposition \cite{cahn1958free}. First we observe from the expression of the pressure (\ref{pressure}) that the Helmholtz free energy has infinitely many minima at $\lambda_0^{(2n)}$ in the spontaneously broken {\cal PT}-regime, where
\begin{equation}
	\lambda_0^{(n)}= - \frac{n^2  \pi ^2 T^2}{N} , \qquad n \in \mathbb{Z},
\end{equation}
denotes the zeros of $p(\lambda)$. Thus we have infinitely many homogeneous equilibrium states in that situation and expect therefore first order phase transitions to occur.

At first we identify the Maxwell line marking the constant pressure of the heterogenous states. We easily find that the $\lambda$-axis constitutes the Maxwell line, i.e. $p_{\text{Max}}=0$, since the areas above and below the isotherms are the same for all temperatures $T$
\begin{equation}
	I^{(n)}:= \!\!\!	\int\limits_ {\lambda_0^{(n-1)}}^{\lambda_0^{(n)}}  \!\!\!  p(\lambda)  d \lambda = T \log \left\{ \frac{\cos \left[ (n-1) \pi \right]- \cosh\frac{\nu}{T}  }{\cos \left[ n \pi \right]-\cosh\frac{\nu}{T} } \right\} , 
\end{equation}
where $ n \in \mathbb{N}$. Thus we have $  I^{(n)}=-   I^{(n+1)}$.

\begin{figure}[h] 
	\begin{minipage}[b]{0.88\textwidth}     
		\includegraphics[width=\textwidth]{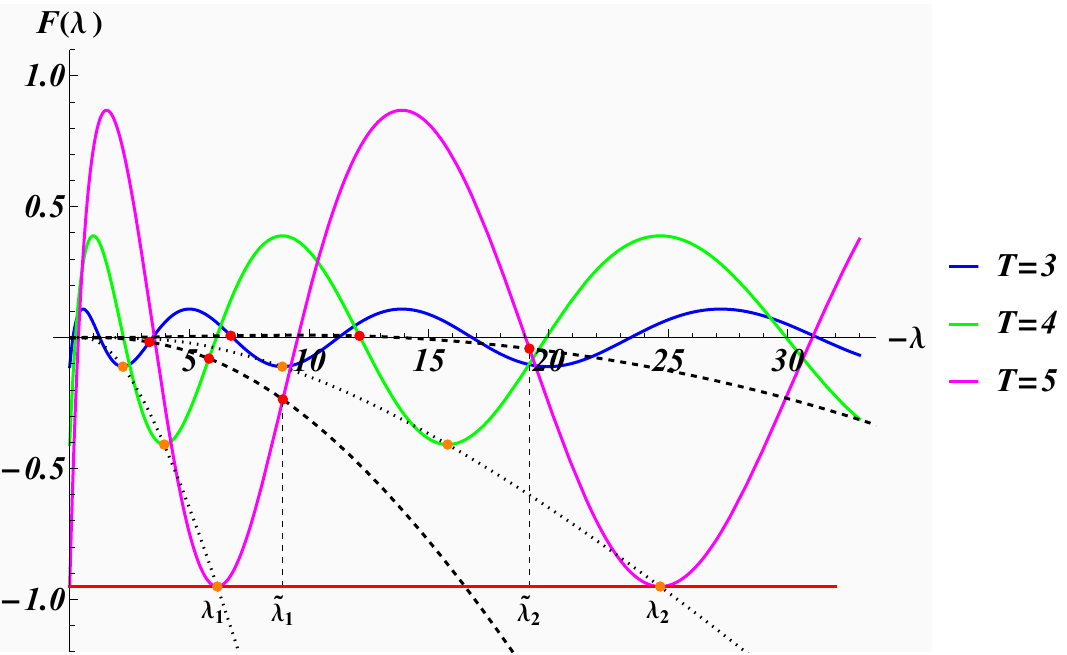}
			\end{minipage}   
	\caption{Helmholtz free energy for the homogeneous states $F(\lambda)$ as functions of the volume $\lambda$ in the spontaneously broken ${\cal PT}$-regime at sample constant temperatures $T$. The Helmholtz free energy for the heterogeneous states $F_{\text{het}}(\lambda) $ is indicated as red line for $T=5$. The binodal region and the spinodal region are bounded by $\lambda_1,\lambda_2$  and the inflection points  $\tilde{\lambda_1},\tilde{\lambda}_2$, respectively. The binodal lines (dotted black) and spinodal lines (dashed lines) are obtained by varying the temperature $T$, intersecting at the critical temperature $T_{\text{crit}}$. }
	\label{spinodal}
\end{figure}

Following \cite{cahn1958free} we next carry out a spinodal decomposition in order to identify the unstable regions and the critical temperature. First we identify the line of heterogeneous states in which the two phases exist together in different proportions 
\begin{equation}
	  n_1 = \frac{\lambda_2-\lambda}{\lambda_2-\lambda_1},  \quad \text{and} \quad n_2= \frac{\lambda-\lambda_1}{\lambda_2-\lambda_1} .
\end{equation}
Denoting by $\lambda_1 = \lambda_0^{(2n)} $ and $\lambda_2= \lambda_0^{(2n+2)}$ two next but one zeros, we can split up the Helmholtz free energy for these states in this binodal region $[\lambda_1 ,\lambda_2 ]$ into a contribution from each of the components according to the so-called lever rule as
\begin{equation}
	F_{\text{het}}(\lambda) = n_1 F(\lambda_1) + n_2 F(\lambda_2).
\end{equation}
Keeping the temperature constant, we obtain from relation (\ref{pressure})
\begin{equation}
	F(\lambda_1)-F(\lambda_2)= - \int_{\lambda_1}^{\lambda_2} p_{1,2}= (\lambda_1 -\lambda_2) p_{1,2} ,
\end{equation}
where $ p_{1,2}$ denotes the pressure on the Maxwell line. Since in our case we have found $ p_{1,2}=p_{\text{Max}}=0$, it follows that $F(\lambda_1)=F(\lambda_2)$ and therefore $F_{\text{het}}(\lambda) =F(\lambda_1)$. In the binodal region $F_{\text{het}}(\lambda)$ is always lower than   the homogeneous free energy $F(\lambda)$, as seen in figure \ref{spinodal} for the particular temperature $T=5$, where $F_{\text{het}}(\lambda) $ is the common tangent to the minima at $F(\lambda_1)$ and $F(\lambda_2)$. In the spinodal region $[\tilde{\lambda_1},\tilde{\lambda}_2]$ the homogeneous states are known to be unstable so that all states will transition to heterogeneous states, whereas in the complement of the binodal region, i.e. $[\lambda_1,\tilde{\lambda}_1]$ and $[\lambda_2,\tilde{\lambda}_2]$, the homogeneous states are known to be metastable, that is stable with respect to infinitesimal perturbations but unstable against finite perturbations. The intersection point of the binodal and spinodal lines is identified as the critical temperature $T_{\text{crit}}$, where the minima and inflection points coincide. In our case we find
\begin{equation}
	T_{\text{crit}} =0 .
\end{equation}
This means in the spontaneously broken ${\cal PT}$-regime we can employ the Maxwell construction for {\em any} temperature. Notice further that these regions repeat periodically as functions of the volume $\lambda$.

Furthermore, for the heterogeneous system we also find that $P(T)=0$ so that by $dS = (\partial p/\partial T)|_\lambda d \lambda$ it follow that the entropy is vanishing throughout the broken ${\cal PT}$-regime. Hence, when viewing the system as consisting of two phases none of the fundamental axioms of thermodynamics are broken.

\section{Conclusion, summary, outlook}
Our main result is that in the low temperature regime of an ensemble build on a non-Hermitian Hamiltonian system in the spontaneously broken ${\cal PT}$-regime three new options exist to connect two values of the entropy at different temperatures that do not manifest in the other regimes: One can connect these points by a) by varying $\nu$ as a function of temperature at constant entropy and $\lambda$, b) by varying the entropy as a function of temperature at constant $\lambda$ and $\nu$, c) by varying the temperature as a function of time at constant entropy, $\lambda$ and $\nu$. The possibility a) can be employed in an ideal Carnot cycle, whereas the possibilities b) and c) allow to set up a new type of cycle along an isochoric path. The new cycle has a better efficiency than the Carnot cycle. The nature of the paths in the new cycle resembles a Stirling cycle  apart from step $2 \rightarrow 3$, but its efficiency is quite different from setting up the latter in an ideal gas. Thus our results appears to contradict a claim made in \cite{gardas2016non} that the classical Carnot bound holds in both ${\cal PT}$-symmetric regimes. Other possibilities to break the bound were previously found for a Hermitian time-dependent harmonic oscillator coupled to a squeezed thermal reservoir \cite{break}. However, as discussed in section IV the system will undergo a  first order phase transition and should be seen as being composed of two phases. As a consequence the entropy vanishes throughout the spontaneously broken ${\cal PT}$ regime. Thus the cycle can not be set up. Since the existence of the new cycle would imply the violation of various fundamental axioms of thermodynamics, one may take this an endorsement for the first order phase transition to take place.  Accepting this mechanism we found for our model that the entropy is vanishing throughout the spontaneously broken ${\cal PT}$-regime.   

Naturally there are several open issues left to explore in future work. We conjecture that the observed features in our model,  i.e., the signs of the heat supply or removal in the steps $i \rightarrow j$ and the efficiency gain when compared to the Carnot cycle for the proposed cycle are universally occurring in the spontaneously broken ${\cal PT}$-regimes of non-Hermitian systems. However, to confirm this one needs to explore more examples and ultimately identify more generic model-independent mechanisms. While these details belong to a process that appears to be unphysical, it seems to be more interesting to explore further the features of the occurring first order phase transition in different type of non-Hermitian systems in the spontaneously broken ${\cal PT}$-regime. In particular the question of whether the entropy in the spontaneously broken ${\cal PT}$-regime is always zero remains to be answered in more generality.

\newif\ifabfull\abfulltrue

\end{document}